\documentclass[12pt,cqg]{iopart}
\pdfoutput=1

\usepackage{amsopn}
\usepackage{iopams}
\usepackage[numbers,square,sort&compress]{natbib}
\usepackage{graphicx}
\usepackage[pdftex]{hyperref}

\newtheorem{theorem}{Theorem}

\DeclareMathOperator*{\wlim}{w-lim}

\begin{document}

\title{How well is our universe described by an FLRW model?}

\author{Stephen R. Green$^1$ and Robert M. Wald$^2$}

\address{$^1$ Department of Physics, University of Guelph\\Guelph, Ontario N1G 2W1, Canada}

\address{$^2$ Enrico Fermi Institute and Department of Physics, University of Chicago\\5640 South Ellis Avenue, Chicago, Illinois 60637, USA}

\eads{\mailto{sgreen04@uoguelph.ca} and \mailto{rmwa@uchicago.edu}}

\begin{abstract}
  Extremely well! In the $\Lambda$CDM model, the spacetime metric,
  $g_{ab}$, of our universe is approximated by an FLRW metric,
  $g_{ab}^{(0)}$, to about 1 part in $10^4$ or better on both large
  and small scales, except in the immediate vicinity of very strong
  field objects, such as black holes. However, derivatives of $g_{ab}$
  are not close to derivatives of $g_{ab}^{(0)}$, so there can be
  significant differences in the behavior of geodesics and huge
  differences in curvature. Consequently, observable quantities in the
  actual universe may differ significantly from the corresponding
  observables in the FLRW model. Nevertheless, as we shall review
  here, we have proven general results showing that---within the
  framework of our approach to treating backreaction---the large
  matter inhomogeneities that occur on small scales cannot produce
  significant effects on large scales, so $g_{ab}^{(0)}$ satisfies
  Einstein's equation with the averaged stress-energy tensor of matter
  as its source. We discuss the flaws in some other approaches that
  have suggested that large backreaction effects may occur. As we also
  will review here, with a suitable ``dictionary,'' Newtonian
  cosmologies provide excellent approximations to cosmological
  solutions to Einstein's equation (with dust and a cosmological
  constant) on all scales. Our results thereby provide strong
  justification for the mathematical consistency and validity of the
  $\Lambda$CDM model within the context of general relativistic
  cosmology.
\end{abstract}

\maketitle

\section{Introduction}

Modern cosmology is based on the assumption of a background FLRW model
with small perturbations. With only six free parameters, the resulting
$\Lambda$CDM model successfully accounts for all cosmological data,
including the CMB and its anisotropies, structure formation and
galactic dynamics, redshift-luminosity relations, and the
nucleosynthesis of light elements. It stands as a truly remarkable
achievement of modern physics.

Nevertheless, there are grounds for questioning the validity---and,
perhaps, even the mathematical consistency---of using the $\Lambda$CDM
model to describe our universe. The observed density contrasts in our
universe are huge on scales of $\sim 1$ Mpc and smaller, and certainly
cannot be described as ``small perturbations'' of an FLRW model in the
usual sense. It might therefore appear that some sort of ``averaging''
over the small scale inhomogeneities must be used in order to even
define the FLRW background model that is supposed to represent our
universe in a $\Lambda$CDM model. If one then succeeds in defining an
averaged FLRW model, how does one know that its dynamics will be
properly described by the dynamical equations applicable to an exact
FLRW model?  Therefore, it is appropriate to ask the question posed as
the title of this paper: ``How well is our universe described by an
FLRW model?''

\begin{figure}
  \centering
  \includegraphics[width=0.7\textwidth]{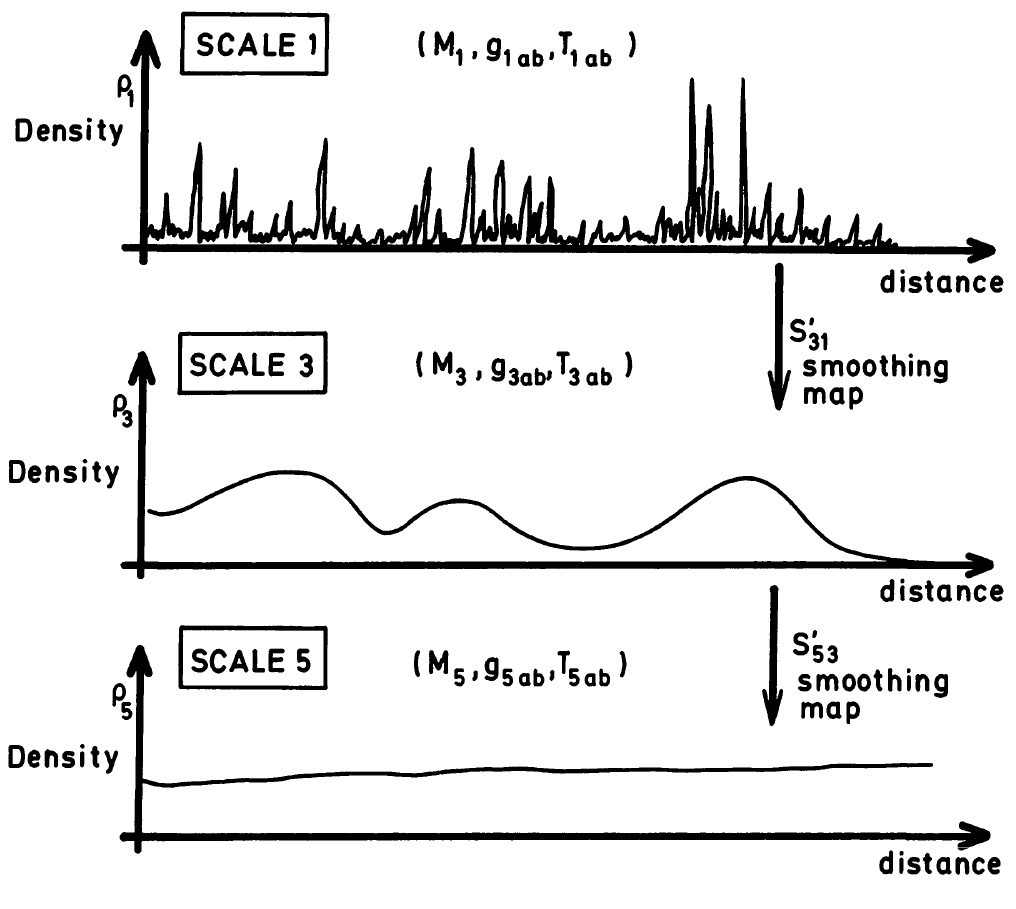}
  \caption{Schematic of a smoothing map, reproduced from
    \cite{Ellis:1984} with kind permission from Springer Science and
    Business Media. Original caption: A comparison of models of the
    same region of the universe on three scales showing different
    amounts of detail.  Scale 1 represents all details down to stars.
    Scale 3 represents all details down to galaxies.  Scale 5
    represents only large scale features.  The models at different
    scales are related by (1) identification of which points at the
    different scales correspond, through maps $S^\ast$; (2) maps $S$
    determining the matter tensor and maps $S'$ determining the metric
    tensor of the smoothed-out representation from the more detailed
    one.  The usual models of cosmology are at Scale 5; but models for
    galaxy formation must be at Scale 3. }
  \label{fig:ellis}
\end{figure}

The issue of how best to approximate our lumpy universe by an FLRW
model---the ``fitting problem''---was discussed in classic papers of
Ellis~\cite{Ellis:1984} and Ellis and Stoeger~\cite{Ellis:1987zz}.
Ellis envisioned defining an operator that would take an inhomogeneous
metric and stress-energy tensor into successively smoother and
smoother versions (see \fref{fig:ellis}).
Given that the Einstein equation holds at the most inhomogeneous scale
(his ``Scale 1''), Ellis raised the very important question of whether
the Einstein equation holds at the level of the large scale averaged metric
and stress-energy tensor.  The averaged Einstein equation is
\begin{equation}
  G_{ab}(g^{(0)})+\Lambda g_{ab}^{(0)} = 8\pi \left( T_{ab}^{(0)} + t^{(0)}_{ab} \right)  .
\label{avgein}
\end{equation}
Here $g_{ab}^{(0)}$ and $T_{ab}^{(0)}$ represent, respectively, the
large scale averaged metric and stress-energy tensor, and
$t_{ab}^{(0)}$ is the average of the (nonlinear) contributions to
Einstein's equation arising from the departures of the actual metric
and stress-energy tensor from $g_{ab}^{(0)}$ and $T_{ab}^{(0)}$.
Thus, $t_{ab}^{(0)}$ represents the {\em backreaction} effects of the
inhomogeneities on the large scale averaged metric, and we will refer
to it as the {\em effective stress-energy tensor} produced by the
inhomogeneities. The issues raised in the previous paragraph then
reduce to the following: (i) How does one define the FLRW quantities
$g_{ab}^{(0)}$ and $T_{ab}^{(0)}$ for the actual universe? (ii) Is
$t_{ab}^{(0)}$ negligibly small?

One of the main points that we wish to emphasize in this article is
that the first question has an extremely simple answer. As we shall
elucidate further below, it is akin to asking how to represent the
surface of a real ball bearing by a perfect sphere. The surface of a
real ball bearing is extremely close to being a perfect
sphere---despite significant imperfections that can be seen under a
microscope---and it is therefore easily represented by a perfect
sphere. Similarly, the actual metric, $g_{ab}$, of our universe is
extremely close to a metric, $g_{ab}^{(0)}$, with FLRW
symmetry---except in the immediate vicinity of very strong field
objects, such as neutron stars and black
holes\footnote{\label{footnote:bh}The large deviation of $g_{ab}$ from
  $g_{ab}^{(0)}$ in the immediate vicinity of strong field objects is
  irrelevant to cosmology.  This is most easily seen by considering a
  new metric, $g_{ab}'$, corresponding to a universe in which each of
  the strong field objects is replaced by an object of the same mass
  but having a radius of $\sim 10^4$ Schwarzschild radii.  Then
  $g_{ab}'$ would be extremely close to $g_{ab}^{(0)}$ everywhere, but
  the cosmological properties of $g_{ab}$ and $g_{ab}'$ would be
  indistinguishable.}---and is easily represented as such. Thus, the
``fitting problem'' for $g_{ab}$ is trivially solved. On the other
hand, as noted above, the actual stress-energy tensor $T_{ab}$ of our
universe is very far from possessing FLRW symmetry. Nevertheless,
since $g_{ab}$ is so close to $g_{ab}^{(0)}$, there is no difficulty
in defining a suitable large scale average of $T_{ab}$, using either
$g_{ab}$ or $g_{ab}^{(0)}$. So, the first question above poses no
difficulty.

However, the second question is less trivial. Although order of
magnitude estimates clearly indicate that $t_{ab}^{(0)}$ should be
negligible for our universe \cite{Ishibashi:2005sj}, one would like to
have a controlled approximation scheme, wherein the dominant
corrections to $g_{ab}^{(0)}$ arising from $t_{ab}^{(0)}$ can be
computed. We have recently developed such an approximation
scheme~\cite{Green:2010qy}, wherein the actual metric $g_{ab}$ is
assumed to be close to a smooth metric $g_{ab}^{(0)}$, but derivatives
of $g_{ab}$ need not be close to derivatives of $g_{ab}^{(0)}$, and
second derivatives (i.e., the curvature) of $g_{ab}$ may have
unbounded fluctuations relative to $g_{ab}^{(0)}$. {\em A priori}, our
approach allows for significant backreaction effects, since the
dominant contribution to $t_{ab}^{(0)}$ comes from quadratic products
of first spacetime derivatives of metric perturbations, which are not
assumed to be small.  Our main result was the proof that---provided
the matter stress-energy tensor satisfies the weak energy
condition---$t_{ab}^{(0)}$ must be traceless and must itself satisfy
the weak energy condition.  Essentially, $t_{ab}^{(0)}$ can be large
only in the presence of significant amounts of gravitational
radiation, in accord with standard intuition. There cannot be any
large backreaction effects arising from matter inhomogeneities. We
will describe our approach and summarize the main results
of~\cite{Green:2010qy} in \sref{sec:framework}.

Even though no large backreaction effects can occur due to deviations
from an FLRW model on small scales, it is still of interest to know
exactly what the corrections due to backreaction are. In particular,
simulations of structure formation are done using Newtonian
cosmological models. Exactly what ``dictionary'' should be used to
translate Newtonian simulations into general relativistic cosmological
models? Exactly how accurate is this dictionary, particularly on large
scales? We have answered these questions in~\cite{Green:2011wc} using
a ``counting scheme'' based upon our approximation method, and we will
summarize our main results in \sref{sec:newtoniancosmology}. We also
include in that section a brief discussion of the degree to which
observable quantities in our universe can differ from an FLRW model.

The above results enable us to provide the following answer to the
question posed in the title of our paper: {\em The metric of our
  universe is extremely well approximated by an FLRW metric that
  satisfies~\eref{avgein} with negligible $t_{ab}^{(0)}$.}
Certain observable quantities involving geodesics (e.g., gravitational
lensing) and curvature (e.g., density contrasts) can differ very
significantly from an FLRW model, but these can be accurately computed
from Newtonian cosmological models using a dictionary of appropriate
accuracy, as described in \sref{sec:newtoniancosmology}.

\begin{figure}
  \centering
  \includegraphics[width=\textwidth]{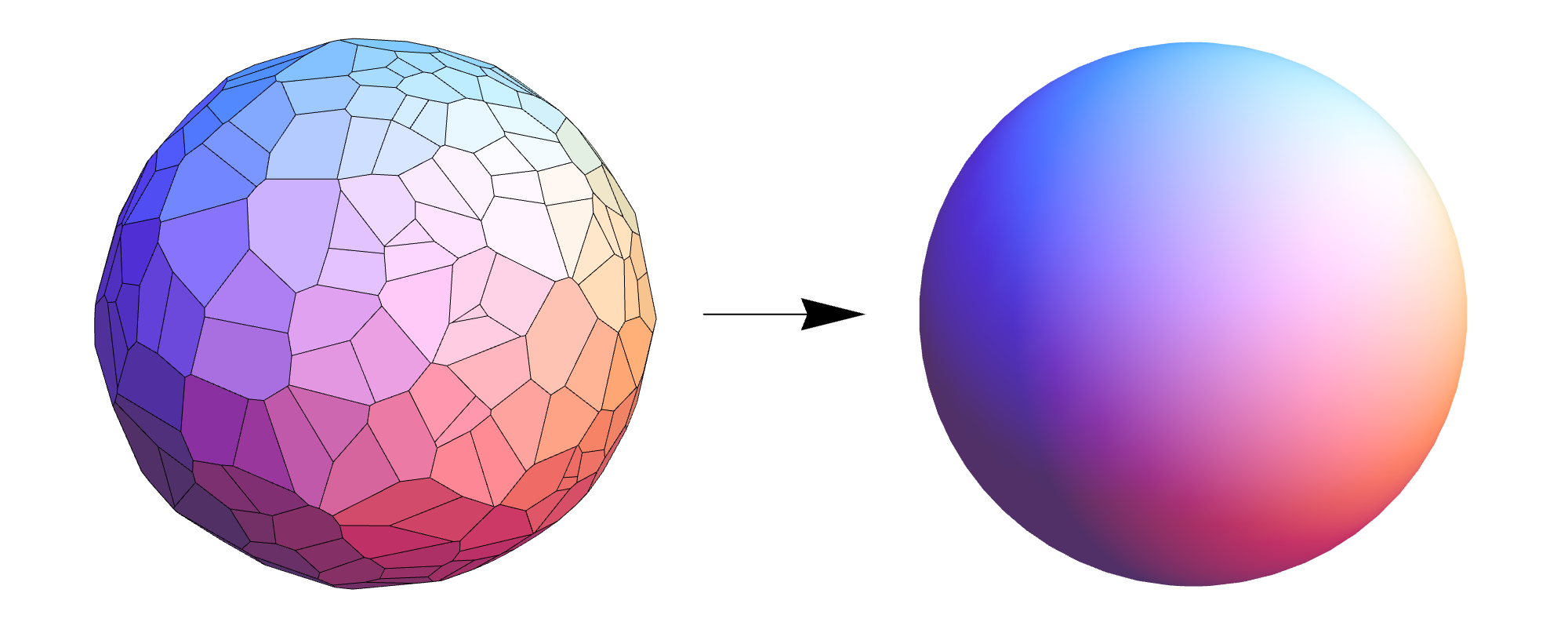}
  \caption{The ``fitting problem'' is analogous to the problem of
    approximating the surface of a ball bearing by a sphere.  Left: A
    300-face convex polyhedron constructed by intersecting the planes
    tangent to 300 randomly chosen points on the surface of a sphere.
    Our ``Sphereland'' is constructed by starting with a similar
    polyhedron with 100 times as many faces, and slightly rounding the
    vertices and edges.  Right: A sphere that solves the ``fitting
    problem'' for the polyhedron (compare with \fref{fig:ellis}).
    Note that we consider a 2-dimensional surface for illustrative
    purposes only.  In higher dimensions, the polygons must be
    replaced by higher dimensional polyhedra and the number of polyhedra
    would have to be corrected, but the construction would be
    otherwise identical.}
  \label{fig:polyhedronsphere}
\end{figure}

Before describing our results in the following sections, we now
elucidate the nature of the ``fitting problem'' for our universe by
exploring more fully the ball bearing analogy alluded to
above. Consider a convex polyhedron with $30,000$ faces, where the
faces are taken to be tangent to randomly chosen points on a unit
sphere. We cannot easily draw such a figure with the required
resolution, so to illustrate the kind of surface we have in mind, we
have drawn such a polyhedron with 300 faces in
\fref{fig:polyhedronsphere}, but it should be kept in mind that the
polyhedron we are considering has a hundred times as many faces as
the one drawn in \fref{fig:polyhedronsphere}. For a polyhedron with
$30,000$ faces of the sort that we are considering, the typical
linear size of each face is $\sim 1/50$ of the radius of the sphere,
and the typical deviation of the polyhedral surface from the unit
sphere is no more than about 1 part in $10^4$. Thus, a solid steel
body with this polyhedral surface would pass inspection as a
semi-precision ball bearing (see~\cite{abbottballs}).

The curvature of our polyhedral surface is concentrated as
$\delta$-functions at the vertices, and the metric of the polyhedron also
fails to be smooth at the edges. We therefore slightly round the
vertices and edges, to produce a smooth surface, with large curvature
near the (former) vertices and nearly flat geometry elsewhere.

In the spirit of George Gamow's discussion of ``Flatland'' in his book
{\it One, Two, Three \ldots Infinity}~\cite{gamow:1947}, we now
consider how creatures making observations on our above constructed
``Sphereland'' would interpret what they experience. Our observers are
assumed to be tiny---much smaller than the typical distance between
the vertices of the original polyhedron. They are blind, and they cannot
move off the surface. Nevertheless, they have ropes that they can pull
tightly between any two points on the surface, and they can measure
rope lengths and angles between ropes extremely accurately. This
allows them to accurately map out the curvature of Sphereland by
measuring the deviation from $180^{\circ}$ of the sum of the angles of
a triangle constructed from taut ropes. When they plot the
curvature, they obtain results that depend upon the size of the
triangle they use, in a manner that qualitatively looks very much like
\fref{fig:ellis}. In particular, when they use triangles that are much
smaller than the typical distance between the vertices, the curvature
fluctuates enormously, usually being very small but occasionally being
extremely large. On the other hand, when they use triangles that are
large compared with the distance between vertices, they obtain results
that are reasonably consistent with spherical geometry. Nevertheless,
even when looking at phenomena on large scales, the observers find
some disturbing anomalies when attempting to model Sphereland by a
perfect sphere. For example, when they measure the angle subtended by
a distant object (by stretching ropes to the ends of the object and
measuring the angle between the ropes), they usually find it to be
somewhat smaller than predicted by the sphere model, but occasionally
they find it to be much larger. They also find that they can sometimes
span more than one taut rope between the same two points, even
though these points are not close to being antipodal in the sphere
model.

As a result of these observations, the observers might be tempted to
conclude that the perfect sphere model provides a reasonably good
description of Sphereland on large scales---although with some
significant deviations---but an extremely poor description of
Sphereland on small scales. However, the actual situation, of course,
is that the metric, $q_{ab}$, of Sphereland is everywhere extremely
close to the metric, $q_{ab}^{(0)}$, of a perfect sphere. This is true
on all scales and does not require any averaging of the Sphereland
metric. Nevertheless, despite the fact that $q_{ab}$ and
$q_{ab}^{(0)}$ are extremely close, their derivatives are not very
close, so, in particular, geodesics determined by $q_{ab}$ can differ
significantly from geodesics determined by $q_{ab}^{(0)}$. This
accounts for the angular size and other anomalies they find in their
large scale measurements. Finally, second derivatives of $q_{ab}$
differ enormously from second derivatives of $q_{ab}^{(0)}$, so the
curvature of $q_{ab}$ differs enormously from that of $q_{ab}^{(0)}$. In
summary, the geometry of Sphereland is described by a metric of the
form $q_{ab}^{(0)} + s_{ab}$, where $q_{ab}^{(0)}$ is the metric of a
perfect sphere and $|s_{ab}| \ll |q_{ab}^{(0)}|$, but first
derivatives of $s_{ab}$ are not small, and second derivatives of
$s_{ab}$ may be enormous.

In an exactly similar manner, the {\em spacetime} metric of our
universe takes the form\footnote{It is fair to ask how we ``know'' the
  facts asserted in this paragraph. As with all scientific
  ``knowledge,'' our beliefs are based on having a small set of simple
  assumptions that account for a vast amount of disparate data in a
  mathematically consistent manner. The $\Lambda$CDM model is based
  upon a simple set of assumptions and successfully accounts for a
  vast amount of disparate data. The results summarized in this
  article confirm that it is mathematically consistent.  Our figure of
  ``1 part in $10^4$'' comes from Newtonian cosmological simulations,
  which yield values of the Newtonian potential in the present
  universe no larger than $\sim 10^{-4}$ (as occurs near the center of
  the richest galaxy clusters); our dictionary (see section
  \ref{sec:newtoniancosmology}) implies that the spacetime metric
  deviations from FLRW are of the same size as the Newtonian
  potential.} $g_{ab} = g_{ab}^{(0)} + \gamma_{ab}$, where
$g_{ab}^{(0)}$ has FLRW symmetry and the components of $\gamma_{ab}$
are extremely small relative to $g_{ab}^{(0)}$---at the level of at
most about 1 part in $10^4$. This is true on all scales---except in
the immediate vicinity of black holes and neutron
stars\footnote{Korzy\'nski~\cite{Korzynski:2013tea} has recently
  extended the Lindquist-Wheeler universe consisting of a lattice of
  black holes~\cite{Lindquist:1957} to irregular black hole
  distributions, and has shown that an FLRW metric arises as a
  continuum limit as the number of black holes goes to infinity.  This
  construction illustrates that even in a universe containing only
  black holes, the metric can be very close to an FLRW metric except
  in the immediate vicinity of the black holes.}---and does not
require any averaging. However, first derivatives of $\gamma_{ab}$ are
not negligibly small, and second derivatives of $\gamma_{ab}$ can be
enormous, in accord with the large density inhomogeneities that we
observe on scales much less than the Hubble radius.

Nevertheless, the fact that $g_{ab}$ is very close to $g_{ab}^{(0)}$
does not imply that backreaction effects of inhomogeneities must be
negligible; as already discussed above, {\it a priori} the quantity
$t_{ab}^{(0)}$ appearing in \eref{avgein} could be large. In the next
section, we will summarize our results showing that, in fact, matter
inhomogeneities cannot produce large backreaction effects. In
\sref{sec:averaging} we will discuss the main flaw of some other
approaches that have suggested that large backreaction effects may
occur. In \sref{sec:newtoniancosmology}, we will describe how to very
accurately translate Newtonian cosmological models into general
relativistic models.

Our notation follows~\cite{Wald:1984}.  Roman letters from the early alphabet,
$\{a,b,c,\dots\}$, denote abstract spacetime indices. Roman letters
from mid-alphabet, $\{i,j,k,\ldots\}$, denote spatial
indices.

\section{Summary of our framework and results}\label{sec:framework}

In this section we briefly review the framework of~\cite{Green:2010qy}
and summarize our main results. Our framework is a generalization to
the non-vacuum case of a framework proposed by
Burnett~\cite{Burnett:1989gp}, which itself is a rigorous version of
Isaacson's approach~\cite{Isaacson:1967zz,Isaacson:1968zza}. Related
approaches that make Newtonian assumptions from the outset are given
in \cite{Futamase:1996fk,Carbone:2004iv,Baumann:2010tm}; see also
\cite{Zalaletdinov:1996aj}.

As indicated in the Introduction, we are interested in describing a
situation where the actual spacetime metric is of the form $g_{ab} =
g_{ab}^{(0)} + \gamma_{ab}$, where $g_{ab}^{(0)}$ has FLRW symmetry
and $\gamma_{ab}$ is small, but where first derivatives of
$\gamma_{ab}$ are not small and second derivatives of $\gamma_{ab}$
may be unboundedly large. The FLRW symmetry of $g_{ab}^{(0)}$ plays no
role in our framework or results, so in this section, $g_{ab}^{(0)}$
may be any smooth metric.

We would like to determine what restrictions Einstein's equation places
on $g_{ab}^{(0)}$. To do so, we could try to proceed by plugging the
metric form $g_{ab} = g_{ab}^{(0)} + \gamma_{ab}$ into Einstein's
equation and collecting terms of the various ``sizes.'' The largest
terms are those that are linear in the second derivatives of
$\gamma_{ab}$ and do not have any other factors of
$\gamma_{ab}$. However, these terms become negligibly small when
averaged. The next largest terms are of two distinct types: (a) terms
made from $g_{ab}^{(0)}$ alone and (b) terms quadratic in
$\gamma_{ab}$ that have a total of two derivatives acting on
$\gamma_{ab}$. When averaged over a spacetime region that is small
compared with the radius of curvature of $g_{ab}^{(0)}$ but large
compared with the variation scale of $\gamma_{ab}$, we will get an
equation of the form \eref{avgein}, where $t_{ab}^{(0)}$ is the
average of the quadratic terms (b). Since all terms in Einstein's
equation contain at most a total of only two derivatives acting on the
metric, all of the cubic and higher order terms in $\gamma_{ab}$
appearing in Einstein's equation become negligible when averaged.

However, if one tries to work with a {\em finite} perturbation $\gamma_{ab}$ in the above manner, it would be difficult to precisely define the ``averaging'' that must be done to obtain to obtain \eref{avgein}, as well as to justify various manipulations of terms. It would also be extremely difficult to obtain any precise, general results on the properties of $t_{ab}^{(0)}$. In order to have precise mathematical rules and obtain precise results, we would like to work with ``infinitesimal'' $\gamma_{ab}$. As in ordinary perturbation theory (see, e.g., section~7.5 of \cite{Wald:1984}), this means working with suitable one-parameter families of metrics $g_{ab} (\lambda)$ such that $[g_{ab} (\lambda) - g_{ab}^{(0)}] \to 0$ as $\lambda \to 0$. However, in order to represent the sort of perturbations that we are interested in, we do {\em not} want to require that spacetime derivatives of $[g_{ab} (\lambda) - g_{ab}^{(0)}]$ go to zero as $\lambda \to 0$. Our precise assumptions on $g_{ab}(\lambda)$ are as follows \cite{Green:2010qy}:
\begin{enumerate}
\item{\label{assumption1} Einstein's equation holds for all $\lambda>0$, i.e., we have 
\begin{equation}
G_{ab}(g(\lambda)) + \Lambda g_{ab}(\lambda) = 8 \pi T_{ab}(\lambda), 
\label{Ee}
\end{equation}
where $T_{ab}(\lambda)$ satisfies the weak energy condition,
i.e., for all $\lambda >0$ we have \mbox{$T_{ab}(\lambda) t^a(\lambda)
  t^b (\lambda) \geq 0$} for all vectors $t^a(\lambda)$ that are
timelike with respect to $g_{ab}(\lambda)$.}
\item{\label{assumption2}There exists a smooth positive function $C_1(x)$ on $M$ such that 
\begin{equation}
|\gamma_{ab}(\lambda,x)| \leq \lambda C_1(x),
\end{equation}}
where $\gamma_{ab} (\lambda,x) \equiv g_{ab}(\lambda,x) - g_{ab}(0,x)$.
\item{\label{assumption3}There exists a smooth positive function $C_2(x)$ on $M$ such that 
\begin{equation}
|\nabla_c \gamma_{ab}(\lambda,x)| \leq C_2(x).
\end{equation}}
\item{\label{assumption4}There exists a smooth tensor field $\mu_{abcdef}$ on $M$ such that 
\begin{equation}
\wlim_{\lambda\to0}
\left[\nabla_a\gamma_{cd}(\lambda)\nabla_b\gamma_{ef}(\lambda) \right] =
\mu_{abcdef},
\label{mu}
\end{equation}
where ``$\wlim$'' denotes the weak limit.}
\end{enumerate}

Here the norms appearing in assumptions (\ref{assumption2}) and (\ref{assumption3}) are taken with respect to any fixed (i.e., $\lambda$-independent) Riemannian metric on spacetime and $\nabla_a$ denotes any fixed derivative operator.
The notion of ``weak limit'', which appears in assumption (\ref{assumption4}),
corresponds roughly to taking a local spacetime average, and then
taking the limit as $\lambda\to0$.  More precisely, $A_{a_1\dots
  a_n}(\lambda)$ is said to {\em converge weakly} to $A_{a_1\dots a_n}^{(0)}$ as
$\lambda\to0$ if and only if, for all smooth tensor fields
$f^{a_1\dots a_n}$ of compact support,
\begin{equation}
  \lim_{\lambda\to0}\int f^{a_1\dots a_n}A_{a_1\dots a_n}(\lambda)=\int f^{a_1\dots a_n}A_{a_1\dots a_n}^{(0)}.
\end{equation}
Assumptions (\ref{assumption2}) and (\ref{assumption3}) capture the notion that $\gamma_{ab}$ is small but its first spacetime derivatives need not be small. Assumption (iv) captures the notion that local spacetime averages of quadratic products of first derivatives of $\gamma_{ab}$ are well behaved.
An explicit example of a one parameter family satisfying assumptions (\ref{assumption1})--(\ref{assumption4}) is given in~\cite{Green:2013yua}.

Assumptions (\ref{assumption1})--(\ref{assumption4}) allow us to
rigorously derive an equation satisfied by $g_{ab}^{(0)}$ of the form
\eref{avgein}, where $t_{ab}^{(0)}$ is given by an explicit formula in
terms of $\mu_{abcdef}$. In \cite{Green:2010qy}, we then proved two
theorems constraining $t_{ab}^{(0)}$:
\begin{theorem}
Given a one-parameter family $g_{ab}(\lambda)$
  satisfying assumptions (\ref{assumption1})--(\ref{assumption4})
  above, the effective stress-energy tensor $t^{(0)}_{ab}$ appearing
  in equation~\eref{avgein} for the background metric $g^{(0)}_{ab}$ is
  traceless,
\begin{equation}
{t^{(0)a}}_a = 0.
\end{equation}
\end{theorem}
\begin{theorem}
  Given a one-parameter family $g_{ab}(\lambda)$ satisfying
  assumptions (\ref{assumption1})--(\ref{assumption4}) above, the
  effective stress-energy tensor $t^{(0)}_{ab}$ appearing in equation~\eref{avgein} for the background metric $g^{(0)}_{ab}$
  satisfies the weak energy condition, i.e.,
\begin{equation}
t^{(0)}_{ab} t^a t^b \geq 0 
\end{equation}
for all $t^a$ that are timelike with respect to $g^{(0)}_{ab}$.
\end{theorem}

In essence, these theorems show that only those small scale metric
inhomogeneities corresponding to gravitational radiation can have a
significant backreaction effect; see \cite{Green:2010qy} for further discussion. In particular, it should be noted that in the case where $g^{(0)}_{ab}$ has FLRW
symmetry, even when short wavelength gravitational radiation is present and backreaction effects are large, the effective stress-energy tensor must be of the form of a
$P=\frac{1}{3}\rho$ fluid, and therefore cannot mimic dark energy.

In~\sref{sec:newtoniancosmology}, we will return to the issue of how
best to describe $g_{ab}$ in cosmology. Specifically, we will consider
how to translate Newtonian cosmologies into general relativistic
cosmologies in a way that yields accurate solutions to Einstein's
equation. However, we first give a brief discussion of some other
approaches that have suggested that backreaction might be large.

\section{Other approaches to backreaction}\label{sec:averaging}

As the results summarized in the previous section have shown, small
scale density inhomogeneities cannot produce large backreaction
effects in cosmology. However, a number of other approaches have come
to the opposite conclusion---or, at least, have suggested the
possibility that the cosmological backreaction effects due to matter
inhomogeneities on small and/or large scales could be large. Some of
these approaches are based upon questionable approximation methods
(see, e.g.,~\cite{Kolb:2005da}) and some are based upon questionable
interpretations of gauge-dependent quantities (see,
e.g.,~\cite{Mukhanov:1996ak,Abramo:1997hu,Unruh:1998ic,Geshnizjani:2002wp,Geshnizjani:2003cn}
with regard to long wavelength perturbations); we will not attempt to
review such approaches here. However, other approaches proceed by
assigning an FLRW metric, $g^{(0)}_{ab}$, to $g_{ab}$ via some
averaging procedure or the matching of some observables. As we shall
now illustrate, the main flaw with such approaches is that the FLRW
metric $g^{(0)}_{ab}$ that one thereby obtains may not be close to
$g_{ab}$---even in cases where $g_{ab}$ is well approximated by some
FLRW metric. Since $g^{(0)}_{ab}$ may be a poor approximation to
$g_{ab}$, there is no reason why $g^{(0)}_{ab}$ need be close to a
solution to Einstein's equation, so one may obtain large backreaction
terms in equations satisfied by $g^{(0)}_{ab}$. Of course, these
apparent backreaction effects are entirely spurious, being generated
by using a poor approximation to $g_{ab}$.

The approach of Buchert~\cite{Buchert:1999er,Buchert:2001sa} provides a good illustration of this point.
In Buchert's approach the
metric is written in a synchronous slicing comoving with the dust
particles,
\begin{equation}\label{eq:synchronousgauge}
  ds^2 = -dt^2 + {}^{(3)}g_{ij} (t) dx^idx^j.
\end{equation}
This is possible provided the dust 4-velocity $u^a$ is irrotational,
which we assume here. The volume, $V_{\mathcal{D}}$, of a comoving spatial region $\mathcal{D}$ on a constant-$t$ slice $\Sigma_t$ is given by
\begin{equation}
V_{\mathcal{D}}(t) = \int_{\mathcal{D}} \,\mathrm{d} \Sigma_t,
\end{equation}
where ${\mathrm{d}}\Sigma_t = \sqrt{{}^{(3)}g(t)} \, \mathrm{d}^3 x$ is the proper volume element on $\Sigma_t$. The averaged ``scale factor,'' $a_{\mathcal{D}}(t)$, is defined as proportional to
$V_{\mathcal{D}}^{1/3}$.
Spatial averages of scalars $\psi$ over $\mathcal{D}$ can then be defined by
\begin{equation}
  \langle \psi \rangle_{\mathcal{D}}(t) \equiv \frac{1}{V_{\mathcal{D}}}\int_{\mathcal{D}} \psi \,\mathrm{d}\Sigma_t  .
\end{equation}
In this way, one can define an averaged density, $\langle\rho\rangle_{\mathcal{D}}(t)$, and averaged scalar curvature, $\langle{}^{(3)}{\mathcal{R}}\rangle_{\mathcal{D}}(t)$. One may then assign an FLRW model, $g^{(0)}_{ab}$, to $g_{ab}$ based upon the values of these averaged quantities. 

Equations governing the dynamics of $g^{(0)}_{ab}$ can be derived
by taking averages of the scalar components of Einstein's equation for $g_{ab}$. One obtains~\cite{Buchert:1999er,Buchert:2001sa}
\begin{eqnarray}
  \label{eq:buchert1}3\left(\frac{\dot{a}_{\mathcal{D}}}{a_{\mathcal{D}}}\right)^2=8\pi\langle\rho\rangle_{\mathcal{D}}-\frac{1}{2}\langle{}^{(3)}{\mathcal{R}}\rangle_{\mathcal{D}}- \frac{1}{2} \mathcal{Q}_{\mathcal{D}},\\
  \label{eq:buchert2}3\frac{\ddot{a}_{\mathcal{D}}}{a_{\mathcal{D}}} = -4\pi\langle{\rho}\rangle_{\mathcal{D}} + \mathcal{Q}_{\mathcal{D}},
\end{eqnarray}
where the quantity
\begin{equation}
  {\mathcal{Q}}_{\mathcal{D}} \equiv \frac{2}{3}\left( \langle \theta^2 \rangle_{\mathcal{D}} - \langle \theta\rangle_{\mathcal{D}}^2\right) - 2 \langle \sigma^2 \rangle_{\mathcal{D}}  .
\end{equation}
characterizes the ``backreaction,'' i.e., the failure of the FLRW spacetime $g^{(0)}_{ab}$, to satisfy Einstein's equation. There is no reason why $\mathcal{Q}_{\mathcal D}$ need be small\footnote{A further
  difficulty with the Buchert approach is that the equations are
  underdetermined, as there is no evolution equation given for
  $\mathcal{Q}_{\mathcal D}$.  However, one is not free to specify
  ${\mathcal{Q}}_{\mathcal D}(t)$ arbitrarily, as the exact (i.e., not averaged)
  metric must satisfy the Einstein equation with dust stress energy, but only two components of this equation have been used to get the Buchert equations.
  See~\cite{Green:2013yua} for further discussion.}, so there can be large backreaction within the Buchert framework. 
  
The large backreaction that can occur in the Buchert framework is a direct consequence of the fact that $g^{(0)}_{ab}$ can be a poor approximation to $g_{ab}$. This is most easily seen in the case where $g_{ab}$ is not globally close to a single FLRW metric\footnote{If the actual metric, $g_{ab}$, of the universe were not globally close to a single FLRW metric, we do not believe that it would be useful to introduce any concept whatsoever of an ``averaged'' FLRW metric $g^{(0)}_{ab}$.}. A nice example of this is provided by a universe that consists of two disconnected dust FLRW universes in different stages of expansion. The Buchert prescription would represent this disconnected universe as a single FLRW universe, which provides a bad approximation to the actual metric everywhere. One can give an example of this sort~\cite{Ishibashi:2005sj} wherein the ``backreaction'' is so large that one obtains acceleration of the representative FLRW universe, even though each of the disconnected components of the actual universe is decelerating. Obviously, this acceleration is not a physical effect but is a spurious artifact of the poor approximation that $g^{(0)}_{ab}$ provides to the actual spacetime metric.

However, the Buchert prescription can also give a poor approximation to the actual spacetime metric $g_{ab}$---and, correspondingly, yield a large backreaction---even when $g_{ab}$ is well approximated by (or even equal to!) {\it some} FLRW metric. An example of this is provided by Minkowski spacetime, which has exact FLRW symmetry. Let $(T,X,Y,Z)$ be global inertial coordinates and consider the hypersurface defined by
\begin{equation}
  T=f(X,Y,Z),
\end{equation}
with 
\begin{equation}
  f(X,Y,Z)=\frac{1}{3N}\sin(NX)\sin(NY)\sin(NZ)  ,
\end{equation}
with $N$ large. This hypersurface is extremely close to the hypersurface $T=0$, but is extremely ``wiggly.'' (However, note that $|\nabla_i f|<1$ everywhere, so the hypersurface is everywhere spacelike.) We choose this hypersurface as the initial hypersurface of a synchronous coordinate system \eref{eq:synchronousgauge}. We take the domain, $\mathcal{D}$, in the initial hypersurface to be a coordinate cube of side $[0,2\pi]$. Since the matter stress-energy vanishes, we obviously have $\langle\rho\rangle_{\mathcal{D}} = 0$. However, we have verified numerically that $\langle{}^{(3)}{\mathcal{R}}\rangle_{\mathcal{D}}$ is enormous for sufficiently large $N$. Thus, although the actual metric, $g_{ab}$, of Minkowski spacetime has exact FLRW symmetry, the Buchert procedure with the above choice of hypersurface instructs us to represent the Minkowski metric with an ``averaged'' FLRW metric $g^{(0)}_{ab}$ that is an extremely poor approximation to $g_{ab}$ on all scales. The backreaction term $\mathcal{Q}_{\mathcal D}$ is correspondingly large. Again, this backreaction is entirely spurious.

One might object to the above example on the grounds that there is no matter present in Minkowski spacetime and we have arbitrarily chosen some artificial hypersurface on which to perform the Buchert construction. However, we could modify this example by adding a tiny sprinkling of dust following geodesics orthogonal to our hypersurface. If the mass density of the dust is sufficiently small, the metric will be essentially unaffected by the dust, but the Buchert procedure will require us to use the above hypersurface and will provide us with an effective FLRW metric that is a very poor approximation to the actual spacetime metric. Of course, the velocity distribution of the dust in this example is not representative of any physically realistic situation. But the example is useful for showing that the Buchert construction depends upon the velocity distribution of the dust even when the dust makes a negligible contribution to the stress-energy. In realistic cosmological situations, the perturbations to the dust velocity will be much larger than the perturbations to the metric (see subsection~\ref{sec:mapping} below), and the comoving synchronous hypersurfaces of the Buchert construction will provide a poor choice for approximating the hypersurfaces with nearly FLRW symmetry. This will lead to a correspondingly large backreaction. We will comment further on this point at the end of subsection~\ref{sec:mapping}.

For similar reasons, large backreaction also appears in the approach
of Clarkson and Umeh~\cite{clarkson2011backreaction}. These authors
obtain expressions for an effective $H$ and $q_0$---and, thereby, a
representative FLRW metric $g^{(0)}_{ab}$---by writing down an
approximate formula for the angular diameter distance $d_{\mathrm{A}}$
as a function of redshift $z$ and performing an average. However,
lensing effects are sensitive to small scale structure since they
depend on derivatives of the metric; as discussed further in
subsection~\ref{sec:observations} below, lensing effects can be very
large even when the deviations of the metric from an FLRW metric are
very small. Consequently, for the power spectra they consider, they
obtain UV divergent expressions for their effective $H$ and $q_0$. The
expressions for backreaction are similarly divergent. Again, these
large backreaction effects arise from the fact that their choice of
FLRW metric, $g^{(0)}_{ab}$, is an extremely poor
approximation\footnote{It may sound odd that an FLRW metric
  $g^{(0)}_{ab}$ that is chosen to be the {\em best fit} to some
  observation made in the nearly FLRW metric, $g_{ab}$, should be an
  extremely poor approximation to $g_{ab}$. However, one might note
  that if our Sphereland observers chose the sphere metric that gave
  the best fit to their observation of, e.g., the average of the {\em
    square} of the curvature, they would get an extremely poor
  approximation to the metric of Sphereland.} to the actual spacetime
metric, $g_{ab}$.

\section{Cosmology with a Newtonianly-perturbed FLRW metric}\label{sec:newtoniancosmology}

As we have argued above, the metric of the universe is well described
by a metric of the form $g_{ab} = g^{(0)}_{ab} + \gamma_{ab}$, where
$g^{(0)}_{ab}$ is an FLRW metric satisfying the usual Einstein
equation with the averaged matter stress-energy as source, and
$\gamma_{ab}$ is ``small''---although its derivatives are not
small. We would like to determine $\gamma_{ab}$ more precisely. It is
clear that $\gamma_{ab}$ does not merely satisfy the linearized
Einstein equation about $g^{(0)}_{ab}$, since nonlinear terms in
derivatives of $\gamma_{ab}$ cannot be ignored. In particular, we
would like to know the leading order effects that these nonlinear
corrections have on large scale behavior, including the expansion rate
of the universe.

We analyzed this issue in~\cite{Green:2010qy} in the context of the completely general framework described in \sref{sec:framework} above. In that reference, we derived an equation satisfied by the ``long wavelength'' part of $\gamma_{ab}$ that takes into account the leading order nonlinear corrections due to the small scale inhomogeneities. However, this equation is extremely complicated, and it is not easy to see what effects it predicts in a general context. To make further progress, it is necessary to put in more assumptions about the nature of our universe. A key observational fact about our universe is that it is locally Newtonian, i.e., as seen by observers ``at rest'' with respect to $g^{(0)}_{ab}$, velocities of nearby objects are nonrelativistic and their dynamics is well described by Newtonian gravity---except, of course, in the immediate vicinity of black holes and neutron stars, which are not relevant for this discussion. Indeed, cosmologists normally simulate structure formation in our universe using only Newtonian gravity---even on scales comparable to the Hubble radius! 

These considerations allow us to reformulate the issue of determining the leading order corrections to $g^{(0)}_{ab}$ as follows~\cite{Green:2011wc}: Consider a cosmological solution to the equations of {\em Newtonian gravity} of the kind produced in numerical simulations by cosmologists. We may ask the following questions: (1) What ``dictionary'' should be used to transcribe this Newtonian cosmology into a general relativistic cosmology? (2) How accurate is this dictionary, i.e., how close does the resulting general relativistic cosmology come to satisfying Einstein's equation? Of course, these questions are not independent: If the dictionary did not produce accurate results, this would indicate that we need a better dictionary. As we shall describe below, there is a straightforward dictionary suggested by the correspondence between linearized Newtonian gravity and linearized general relativity. This dictionary is quite satisfactory, but it can be further improved~\cite{Green:2011wc} to take into account post-Newtonian effects at small scales as well as the leading order large scale effects produced by the nonlinear effects of inhomogeneities. The resulting improved dictionary should yield an extremely accurate general relativistic description of our universe, and, in particular, it provides the leading order backreaction effects at large scales produced by small scale inhomogeneities.

We will now review Newtonian cosmology. We will then present a dictionary that
maps Newtonian solutions into general relativistic cosmologies and discuss its accuracy.  Finally we discuss some effects that these
inhomogeneities have on light propagation and observables.

\subsection{Newtonian cosmology}

It is a remarkable fact that the evolution equations for a homogeneous and isotropic cosmology with dust matter and a cosmological constant are precisely the same
in Newtonian gravity and general relativity.  To see this, begin with
the equations of Newtonian gravity with a cosmological constant,
\begin{eqnarray}
  \label{eq:poisson}\partial^i\partial_i\phi+\Lambda=4\pi\rho,\\
  \label{eq:masscons}\partial_t\rho+\partial_i(\rho v^i)=0,\\
  \label{eq:euler}\partial_t(\rho v^i)+\partial_j(\rho v^i v^j)=-\rho\partial^i\phi.
\end{eqnarray}
Plugging in the cosmological ansatz, $\rho=\rho_0(t)$, $v^i=H(t)x^i$,
and requiring solutions to be non-singular, it can be
seen~\cite{Peebles1980} that \eref{eq:poisson}--\eref{eq:euler}
reduce to the Friedmann equations for a dust cosmology,
\begin{eqnarray}
  \label{eq:friedmann1}\frac{dH}{dt}+H^2=-\frac{4\pi}{3}\rho_0+\frac{\Lambda}{3},\\
  \label{eq:friedmann2}H^2=\frac{8\pi}{3}\rho_0+\frac{\Lambda}{3}-\frac{k}{a^2}.
\end{eqnarray}
Here $a$ is the radius of any comoving ball, so that
\begin{equation}
  H=\frac{1}{a}\frac{da}{dt}.
\end{equation}
The constant $k$ is a constant of integration, and the density $\rho_0$ varies as $1/a^3$.  Thus, there is an obvious correspondence 
between Newtonian and relativistic dust cosmologies in the homogeneous and isotropic case, which takes exact solutions to exact solutions. In the following, we restrict consideration to the $k=0$ case. 

It is useful to rewrite the exact Newtonian equations relative to a background
solution and to transform to a comoving coordinate system of the
background. We obtain~\cite{Green:2011wc}
\begin{eqnarray}
  \label{eq:poisson1}\partial^i\partial_i\psi_{\mathrm{N}}=4\pi a^2\rho_0\delta_{\mathrm{N}},\\
  \label{eq:masscons1}\dot{\delta}_{\mathrm{N}}+\partial_i((1+\delta_{\mathrm{N}})v_{\mathrm{N}}^i)=0,\\
  \label{eq:euler1}\dot{v}_{\mathrm{N}}^i+v_{\mathrm{N}}^j\partial_jv_{\mathrm{N}}^i+\frac{\dot{a}}{a}v_{\mathrm{N}}^i=-\partial^i\psi_{\mathrm{N}},
\end{eqnarray}
where $v_{\mathrm{N}}^i$ is the velocity relative to the ``Hubble flow,'' $\psi_{\mathrm{N}}$ is the Newtonian potential relative to the background, and $\delta_{\mathrm{N}}$ is the fractional density perturbation. We have inserted the subscript ``N'' on
these expressions in order to be able to clearly distinguish them from the corresponding general relativistic quantities that we will consider below. Equations \eref{eq:poisson1}--\eref{eq:euler1} (or variants describing
discrete matter sources) form the basis of numerical simulations
exploring structure formation in cosmology. 

\subsection{Mapping to general relativity}\label{sec:mapping}

It is perhaps an even more remarkable fact that the exact
correspondence between homogeneous, isotropic Newtonian and general
relativistic dust cosmologies also extends to linearized solutions in the
$k=0$ case, i.e., linearized solutions of
\eref{eq:poisson1}--\eref{eq:euler1} correspond exactly to (scalar and
vector sector) solutions of the $k=0$ general relativistic
perturbation equations~\cite{Bardeen:1980kt} when an appropriate
``dictionary'' is used. For scalar perturbations, this is most easily
seen by noting that, when written in terms of Bardeen's gauge
invariant variables, the Einstein equation with dust matter,
linearized off of FLRW, is identical to the linearized form of
\eref{eq:poisson1}--\eref{eq:euler1}, with the identifications
\begin{eqnarray}
  \label{eq:Bardeen1}\psi_{\mathrm{N}} \longleftrightarrow \Phi_{\mathrm{A}}=-\Phi_{\mathrm{H}},\\
  \label{eq:Bardeen2}v_{\mathrm{N}}^i \longleftrightarrow v_{\mathrm{s}}^i,\\
  \label{eq:Bardeen3}\delta_{\mathrm{N}} \longleftrightarrow \epsilon_{\mathrm{m}}  ,
\end{eqnarray}
where we have used Bardeen's notation for the gauge invariant variables $\Phi_{\mathrm{A}}$, $\Phi_{\mathrm{H}}$, $v_{\mathrm{s}}^i$, and $\epsilon_{\mathrm{m}}$.
A similar correspondence holds in the vector sector, although there it is necessary to also solve a Poisson-type
equation to obtain the vector-type metric perturbation~\cite{Green:2011wc}. The
spacetime metric and stress-energy tensor may be straightforwardly
reconstructed in any gauge from the gauge invariant variables.

While the gauge invariant Bardeen variables are extremely convenient
to use in linear perturbation theory, there are no analogs of them
beyond linear order. Thus, in order to attempt to extend the above
exact Newtonian-relativistic correspondence for linearized solutions
to an approximate Newtonian-relativistic correspondence for exact
solutions, we must make a choice of gauge. We choose to work in
``longitudinal gauge,'' where the metric takes the form
\begin{equation}
  \label{eq:metric-longitudinal}ds^2=a^2(\tau)\left\{-(1+2A)d\tau^2 - 2B_i dx^id\tau + [(1+2H_{\mathrm{L}})\delta_{ij}+h_{ij}]dx^idx^j\right\},
\end{equation}
where $\partial^i B_i = \partial^j h_{ij} = h^i_{\phantom{i}i} =
0$. As argued in~\cite{Green:2011wc}, it should be possible to make
this gauge choice for metrics that are sufficiently close to FLRW
metrics. We will comment further on this gauge choice below.

We could obtain a dictionary by simply transcribing the linearized
dictionary to the nonlinear regime, replacing the Bardeen variables in
the linearized correspondence with their metric component counterparts
in longitudinal gauge.  However, we can easily improve this dictionary
by slightly modifying the definitions of $v^i$ and $B^i$ so as to
obtain consistency with the nonlinear momentum constraint at small
scales.  We obtain~\cite{Green:2011wc}
\begin{eqnarray}
  \label{eq:dict1}A=-H_L=\psi_{\mathrm{N}},\\
  \label{eq:dict2}(1+\delta_{\mathrm{N}})v^i=(1+\delta_{\mathrm{N}})(v_{\mathrm{N}}^i+B^i)-\overline{(1+\delta_{\mathrm{N}})v_{\mathrm{N}}^i},\\
  \label{eq:dict3}\delta=\delta_{\mathrm{N}}-\frac{3}{4\pi\rho_0a^2}\left[\left(\frac{\dot{a}}{a}\right)^2\psi_{\mathrm{N}}+\frac{\dot{a}}{a}\dot{\psi}_{\mathrm{N}}\right],\\
\label{eq:dict4}\partial^j\partial_j B^i = -16\pi\rho_0a^2\left.\left((1+\delta_{\mathrm{N}})v_{\mathrm{N}}^i-\overline{(1+\delta_{\mathrm{N}})v_{\mathrm{N}}^i}\right)\right|_{\mathrm{v}},\\
  \label{eq:dict5}h_{ij}=0.
\end{eqnarray}
In these expressions, the overline denotes an average over all of
space, while the subscript ``v'' denotes the ``vector part''.  Note
that it is necessary to solve the Poisson equation~\eref{eq:dict4} to
obtain $B^i$, but this is a straightforward procedure that could be
incorporated into numerical simulations.  The quantity $B^i$ and the
corrections to $v^i$ will normally be small.  If we neglect these
terms, we obtain a continuum version of the dictionary developed
in~\cite{Chisari:2011iq}.  We note, however, that very recent efforts
to incorporate $B^i$ into Newtonian simulations have suggested that it
may be possible to measure the frame dragging effects of
$B^i$~\cite{Bruni:2013mua,Adamek:2013wja}.

If we start with an exact Newtonian cosmology and translate it into a
general relativistic cosmology using the above dictionary, we wish to
know how well Einstein's equation is satisfied as well as the
corrections to the dictionary that would be needed to make Einstein's
equation hold to greater accuracy. In order to proceed, we need to
have an {\it a priori} notion of the ``size'' of the Newtonian
quantities. To assign such orders, we break up each quantity into its
(``long wavelength'') averaged part and its (``short wavelength'')
remainder. We introduce a ``small parameter'' $\epsilon$, and take the
long wavelength parts of $\psi_{\mathrm{N}}$, $v_{\mathrm{N}}$, and
$\delta_{\mathrm{N}}$ to be $\Or(\epsilon)$. We also take all space
and time derivatives of the long wavelength parts of these quantities
to be $\Or(\epsilon)$. On small scales, where nonlinear motion is
important, our assignment of ``sizes'' is based on usual
post-Newtonian counting.  We take $\psi_{\mathrm{N}}=\Or(\epsilon)$
and $v_{\mathrm{N}}=\Or(\epsilon^{1/2})$. Each spatial derivative is
assumed to contribute an inverse power of $\epsilon$, $\partial_i \sim
\Or(\epsilon^{-1})$, and each time derivative is assumed to contribute
an inverse half-power of $\epsilon$, $\partial_0\sim
\Or(\epsilon^{-1/2})$. Consistent with the Newtonian equation
\eref{eq:poisson1}, we then have
$\delta_{\mathrm{N}}=\Or(\epsilon^{-1})$.  Oliynyk has proven
existence of a wide class of {\em one-parameter families} of
inhomogeneous solutions to the Einstein equation with this
behavior~\cite{Oliynyk:2013jqa,Oliynyk:2014ufa}.

With this counting scheme in hand, we can check how well the Einstein
equation is satisfied by the relativistic cosmology given by our
dictionary. {\it A priori}, the long wavelength part of Einstein's
equation contains terms at most $\Or(\epsilon)$, whereas the short
wavelength part of Einstein's equation contains terms that are
$\Or(\epsilon^{-1})$. We find that, with the above dictionary, the
long wavelength part fails to hold at $\Or(\epsilon)$---due to the
appearance of such terms as the average of quadratic products of
derivatives of the short wavelength part of
$\psi_{\mathrm{N}}$---whereas the short wavelength part of Einstein's
equation fails at $\Or(1)$. However, we showed in~\cite{Green:2011wc}
how to improve our dictionary so that the long wavelength part of
Einstein's equation holds at $\Or(\epsilon)$, and the short wavelength
part of Einstein's equation holds at $\Or(1)$. The short wavelength
corrections are post-Newtonian corrections and are negligible for any
present-day observations.  The long wavelength corrections essentially
correspond to including as source terms in Einstein's equation the
effects of the energy and stresses of the Newtonian potential and the
kinetic motions. {\em These provide the leading order backreaction
  effects on large scales of the small scale inhomogeneities.}
Although these corrections are formally $\Or(\epsilon)$ in our
counting scheme, they are, in fact, negligible\footnote{In fact, the
  next order correction to the pressure vanishes completely when
  averaging virialized systems~\cite{Baumann:2010tm}; this was
  recently demonstrated numerically~\cite{Adamek:2014gva}.  The
  correction to the energy density corresponds to the inclusion of
  kinetic energy and Newtonian binding energy.}. Thus, while it is
very comforting to know that the improved ``Oxford dictionary''
of~\cite{Green:2011wc} exists, the
dictionary~\eref{eq:dict1}--\eref{eq:dict5} should provide an
excellent description of our universe. The smallness of Newtonian
perturbations guarantees that the universe is well-described by an
FLRW model, even though the stress-energy perturbations are very large
on small scales.


Finally, we make a comment on our choice of metric
form~\eref{eq:metric-longitudinal}. It is extremely important that we
did {\em not} choose a metric form---such as the comoving synchronous
form~\eref{eq:synchronousgauge}---that is tied to the matter motion,
i.e., geodesics of the actual spacetime. This is because the geodesics
of the actual spacetime differ significantly from the geodesics of the
background spacetime, so the metric components in, e.g., synchronous
coordinates can be very different even when the metrics are actually
very close. In the present instance, this phenomenon is manifested as
follows: Even at the linear level, when the
dictionary~\eref{eq:Bardeen1}--\eref{eq:Bardeen3} is expressed in a
particular gauge, there will be some ``mixing'' between the velocity,
the potential, and the density.  For example, in the longitudinal
gauge, the relativistic density perturbation \eref{eq:dict3} involves
$\psi_{\mathrm{N}}$. However, in the longitudinal gauge, the
dictionary provides metric perturbation variables that consistently
remain $\Or(\epsilon)$ in our counting scheme. By contrast, in the
comoving synchronous gauge, metric components pick up contributions
from $v_{\mathrm{N}}$. These cause the metric perturbation to become
$\Or(1)$ on a timescale of order $\epsilon^{1/2}$, thereby potentially
leading to large, spurious backreaction effects. This is closely
related to the example given near the end of \sref{sec:averaging}.

\subsection{Effects on observables}\label{sec:observations}

As we have emphasized throughout this article, the spacetime metric, $g_{ab}$, of our universe is very close to an FLRW metric, $g_{ab}^{(0)}$. However, most things that we {\em observe} about the universe involve electromagnetic radiation, which propagates on null geodesics. However, the 
geodesics of the actual metric $g_{ab}$ are {\em not} necessarily very close to corresponding geodesics of  $g_{ab}^{(0)}$.
Moreover, the time evolution of the shear and convergence of a bundle of
geodesics depends on the Riemann curvature~\cite{Wald:1984} (i.e., second
derivatives of the metric), which can be very large.  As a result,
observables in the actual universe described by $g_{ab}$ can differ
considerably from observables computed in the FLRW background
spacetime $g_{ab}^{(0)}$.

The redshift--luminosity relation provides a good example of the kinds
of differences that can occur for observables, even when $g_{ab}$ and
$g_{ab}^{(0)}$ are very close. In an FLRW spacetime, the Weyl
curvature vanishes and the Ricci curvature is uniform. However, in the
actual universe, most of the matter is clumped into structures such as
galaxies. If a beam of light passes through a clump of matter, it will
encounter high Ricci curvature and experience large convergence. If it
passes near---but not through---a clump of matter, it will encounter
only Weyl curvature, producing shear and, eventually, convergence. If
it does not pass near any clumps of matter, it will not experience any
significant curvature at all. Thus, if the actual universe is
sufficiently lumpy, then the light beams from most sources will
encounter little curvature and will be dimmed compared with the FLRW
spacetime~\cite{Dyer:1972distance}. However, occasionally, the beams
will pass through strong curvature, and the sources will be greatly
magnified and/or multiply imaged. Thus, large and highly non-Gaussian
deviations in the apparent luminosity of standard candle high redshift
sources will occur~\cite{Holz:1997ic,holz1998lensing}.

By flux conservation, the {\em average} of the apparent luminosity of
sources (including multiple images) must match the FLRW value to a
high degree of accuracy. Thus, if one were to plot the {\em mean
  apparent luminosity} of standard candle sources versus redshift, one
should obtain good agreement with the underlying FLRW model. However,
if one were to plot the mean of a nonlinear function of
luminosity---such as the luminosity distance---versus redshift, one
could get significant departures from the underlying FLRW values at
high redshift, on account of the large, non-Gaussian
fluctuations. This illustrates the
point that, as a general matter of principle, one should {\em not}
start at the level of observables and then try to design some
effective FLRW metric that captures the behavior of these
observables. Rather, one must proceed as follows:

\begin{enumerate}
\item Start with a background FLRW metric that is believed to be close
  to the actual metric of the universe (except, of course, in
  the immediate vicinity of strong-field objects such as black holes).
\item Determine the corrections to the FLRW metric that describe
  inhomogeneities.  To an excellent approximation, these are the
  (small) Newtonian corrections described in the previous subsection.
\item Calculate observables in the full metric (background plus
  perturbations).
\end{enumerate}

The $\Lambda$CDM model has been remarkably successful in carrying out
this strategy and thereby making predictions in excellent agreement
with all cosmological observations. As we have reviewed in this
article, this model is mathematically consistent and fully in accord
with general relativity. Unless some observational discrepancy arises
in the future, the $\Lambda$CDM model deserves to be ranked among the
great achievements of modern science.

\ack

This research was supported in part by NSF
grant PHY 12-02718 to the University of Chicago, and by NSERC. SRG is
supported by a CITA National Fellowship at the University of Guelph,
and he thanks the Perimeter Institute for hospitality. Research at
Perimeter Institute is supported through Industry Canada and by the
Province of Ontario through the Ministry of Research \& Innovation.

\bibliography{mybib.bib}

\end{document}